\documentclass[aps,twocolumn]{revtex4}
\usepackage{graphicx}
\usepackage{amsmath}
\begin{document}
\newcommand{\rr}{{\bf r}}
\newcommand{\kk}{{\bf k}}
\newcommand{\pp}{{\bf p}}
\newcommand{\qq}{{\bf q}}
\newcommand{\Q}{{\bf Q}}
\newcommand{\BSCCO}{{Bi$_2$Sr$_2$CaCu$_2$O$_8$ }}
\newcommand{\YBCO}{{YBa$_2$Cu$_3$O$_{7-\delta}$ }}
\def\k{{\bf k}}
\def\r{{\bf r}}
\def\R{{\bf R}}
\def\p{{\bf p}}
\def\D{{\cal D}}

\title{Quantum interference in nested $d$-wave superconductors:
a real-space perspective} \author{W. A. Atkinson$^1$, P. J.
Hirschfeld$^{2,3}$, and Lingyin Zhu$^{2,3}$\\~}

\affiliation{$^1$Department of Physics, Southern Illinois
University, Carbondale IL 62901-4401\\
$^2$Department of Physics, University of Florida,
Gainesville FL 32611\\
$^{3}$Center for
Electronic Correlations and Magnetism, EP6, Univ.  Augsburg,
Augsburg Germany}
\date{\today}

\begin{abstract}
We study the local density of states around potential scatterers
in $d$-wave superconductors,
and show that quantum interference between impurity states is not
negligible for experimentally relevant impurity concentrations.  The
two impurity model is used as a paradigm to understand these effects
analytically and in interpreting numerical solutions of the
Bogoliubov-de Gennes equations on fully disordered systems. We focus
primarily on the globally particle-hole symmetric model which has been
the subject of considerable controversy, and give evidence that a
zero-energy delta function exists in the DOS.  The anomalous spectral
weight at zero energy is seen to arise from resonant impurity states
belonging to a particular sublattice, exactly as in the 2-impurity
version of this model. We discuss the implications of these findings
for realistic models of the cuprates.

\end{abstract}


\maketitle

\section{Introduction}

Improvements in high-resolution scanning tunneling microscopy (STM)
applied to superconductors
\cite{yazdani,davisnative,davisZn,cren,davisinhom1,davisinhom2,Kapitulnik1,Kapitulnik2,Hoffman1}
have raised the prospect of obtaining completely new kinds of local
information about the cuprate materials, which may bear on the origins
of the high-temperature superconductivity itself.  Interpretation of
these experiments is understood to be a delicate matter, but until now
has been undertaken at only the naivest levels for want of theoretical
tools for studying the local properties of strongly correlated
systems.  As an example, one may consider the discovery of subgap
impurity resonances at low temperatures in the superconducting state
by STM\cite{yazdani,davisnative,davisZn}: while comparisons of STM
data on disordered BSCCO-2212 with the simplest calculations of a
single potential scatterer in a $d$-wave
superconductor\cite{scalapino,Balatsky,otherlocaldos} were understood
early on to be only approximately successful, it was immediately
proposed\cite{vojta,Zhufilter,martinbalatsky} that more complicated
(but still local) 1-impurity Hamiltonians or STM tunneling matrix
elements could resolve the discrepancies.  Only recently has it been
pointed out that quantum interference of impurity states might make it
difficult to observe true 1-impurity properties at
all\cite{Morr,zhutwoimpurity}. In order for STM to fulfill its
promise, it is vital to understand the extent to which long-range
quantum interference due to disorder influences ostensibly local
properties.

The problem of low-energy $d$-wave quasiparticle excitations in the
cuprates in the presence of disorder is still unsolved (for a review,
see \cite{HAreview}). Traditionally, it has been assumed that the
appropriate disorder potential is some random distribution of
short-range (and possibly magnetic) scatterers.  More recently, there
has been a gradual recognition that nanoscale spatial inhomogeneities
are frequently, and possibly always, present in HTSC.
\cite{cren,davisinhom1,Kapitulnik1,Hoffman1} In most current theories,
disorder is treated in the so-called self-consistent $T$-matrix
approximation (SCTMA) which makes predictions for macroscopic
properties of disordered systems.  The SCTMA predicts, for example, a
constant residual Fermi level density of quasiparticle states
$\rho(0)$, which should dominate the low-energy transport over an
energy range $\gamma$ referred to as the ``impurity bandwidth", in
analogy to similar phenomena in semiconductors.  Transport and
thermodynamic measurements on the cuprates appear to support
qualitatively the predictions of this simple approach though there are
lingering quantititative differences which require
resolution\cite{Hussey}.  The SCTMA neglects ``crossing diagrams"
corresponding to self-retracing scattering paths in real space, and
attempts to go beyond the SCTMA have produced a variety of strongly
model-dependent results for the density of states (DOS), many of which
do not support the idea of an impurity band (constant DOS energy
range) at all. In these nonperturbative calculations, the asymptotic
limit $\rho(0)$ may vanish\cite{tsvelik,fisher,atkinsonops}, saturate
at a finite value\cite{ziegler} or
diverge\cite{pepinlee,mudry,otherdivergence} depending on the symmetry
of the Hamiltonian\cite{atkinsondetails,yashenkin}.  We also note a
recent semi-classical treatment of extended impurities suggesting a
divergent density of states at the Fermi level\cite{goldbart}.

In this paper, we perform simple, exact calculations of the
interference of two impurities in a $d$-wave superconductor, and
compare to numerical calculations for many-impurity systems, in
order to investigate the formation of the impurity band.
Spatial fluctuations in the local DOS, which become quite
complicated as a result of interference between impurities,
contain information about both the SCTMA impurity band and about
the quantum interference processes responsible for weak
localization physics. For purposes of this paper, it is useful to
make a distinction between quantum interference associated with
weak localization and local interference patterns seen, for
example, in STM experiments.
We restrict ourselves in this initial work to a half-filled,
tight-binding band with infinite potential scatterers.  This model
has nesting symmetries which distinguish it from the cuprate
superconductors, but is nevertheless interesting from two points
of view, that transparent analytical results for some properties
can be obtained, and that the character of the divergence of the
density of states near half-filling is
controversial\cite{HAreview}. The two-impurity problem is the
simplest problem which includes the interference processes which
lead to the formation of the impurity band, as well as processes
which lead to weak localization.

Early work on the two-impurity problem in a $d$-wave superconductor
was numerical in nature and focussed on the local density of states
(LDOS), exhibiting unusual local interference patterns which depended
on the orientation of the vector $\bf R$ separating the two
impurities.\cite{Miyake} More recently, the relation to impurity band
formation was discussed\cite{micheluchi} and predictions were made for
STM experiments\cite{Morr,zhutwoimpurity}, assuming that "sufficiently
isolated" two impurity configurations could be identified.  In
reference \cite{zhutwoimpurity}, the bound state wavefunctions of the
two-impurity system were identified and classified.  By analogy with
the molecule problem in quantum mechanics, one expects that the
single-impurity resonance energies split as the impurities are brought
together, and that the wavefunctions are formed from symmetric and
antisymmetric combinations of the isolated impurity wavefunctions.  In
fact, because of the particle-hole and fourfold rotational symmetries
of the superconducting state, the situation is more complicated, with
the effective overlap depending on $\bf R$ . Indeed, it has been shown
that for many pair configurations, the density of states does not
consist of four well-defined resonances\cite{Morr,zhutwoimpurity}.

The interference between impurities persists up to large impurity
separations. In Ref. \cite{zhutwoimpurity} it was noted that two
impurities with ${\bf R}\parallel (110)$ could cause splittings
comparable to the original resonance energy for $R$ of many tens
of lattice spacings.  The spatial LDOS maps  are therefore very
different from superposed single-impurity maps, and one may ask
the question whether this distinction persists in the case of many
impurities. That is, is it to be expected at experimental impurity
concentrations that a resonance found by STM really corresponds to
an isolated impurity whose LDOS is predictable within a simple
1-impurity model\cite{zhutwoimpurity}? Alternatively, are
interference effects omnipresent, destroying expected 1-impurity
resonances and leading to new, long range LDOS patterns which
require a many-impurity interpretation? If the latter scenario is
realized, how can it be that STM experiments seem to see such
similar spectra on or near impurities embedded in very different
local disorder environments? We resolve these questions below by
arguing that in the generic case the individual many-impurity
eigenstates are highly distorted from mere superpositions of
1-impurity LDOS patterns, but that STM measurements tend to
average over many such eigenstates, cancelling some of the
long-range effects of interference. Exceptions are very
low energy states of the {\it nested} d-wave superconductor, which
experience symmetry-driven level repulsion effects which prevent
such cancellations.  These considerations lead to a picture where,
with the exception of the zero energy states, the local impurity
resonances appear homogeneously broadened to any probe which
averages over a macroscopic energy window.
This result has important
consequences for the interpretation of STM spectra.  It means that,
while the position of a peak and the crude local LDOS pattern at an
energy near the peak may indeed qualitatively reflect one-impurity
properties, e.g. the strength of the 1-impurity potential $V_0$, the
widths of spectral features measured at any site will reflect the
impurity bandwidth $\gamma$ characteristic of the disordered system as
a whole.

\begin{figure}
\begin{center}
\includegraphics[width=\columnwidth]{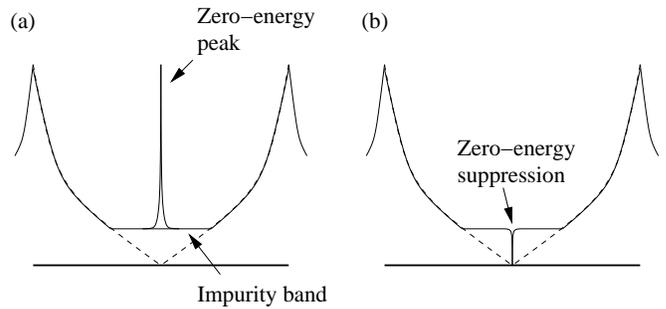}
\caption{Schematic figure of the many-impurity DOS (a) in the
unitary limit of the half-filled band and (b) in the unitary limit
of a generic band. The plateau in the impurity band is characterized by
a nearly constant density of states $\rho_0$.  The zero-energy suppression
in (b) is discussed in Refs.\ \protect \cite{fisher,atkinsondetails}
}
\label{figschematic}
\end{center}
\end{figure}

The second goal of this paper is to investigate the divergence in
the total density of states in the completely nested model from a
local point of view, applying what we can learn about the
2-impurity system. In contrast to the prediction of a residual
$\rho(0)$ by the SCTMA, P\'epin and Lee (PL)\cite{pepinlee} found
that, for an infinite scattering potential, the disorder-averaged
density of states should diverge at the Fermi level (taken to be
the zero of energy here) as $\rho(\omega) \approx
n_i/\omega\log^2\omega$ where $n_i$ is the density of impurities.
The schematic picture of the total DOS in this case is exhibited
in Fig. 1a), to be contrasted with the more generic case expected
in the absence of the nesting symmetry (1b). One surprising aspect
of the PL result for the $N$-impurity $T$-matrix is that it is
essentially $n_i$ times the single-impurity result. Upon closer
inspection, however, their result is not directly tied to the
one-impurity resonance at the Fermi-level, but is the result of
interference between distant impurities. Numerical calculations
showed that the divergence arises because of a global
particle-hole symmetry\cite{atkinsondetails} which is particular
to the tight-binding model at half-filling. It was later shown
that this nesting leads to a novel diffusion mode\cite{yashenkin}
producing a positive logarithmic correction to the DOS. This
general structure of the divergence has also been found by Mudry
and Chamon\cite{mudry} and numerical calculations\cite{Zhu} seem
to confirm it, although in both cases the strength of the
divergence could not be verified. The situation is not settled,
however, and other recent field-theoretical
approaches\cite{otherdivergence} find a different form for the
divergence which is reminiscent of the half-filled normal metal.
The investigations of the 2-- and many-impurity problem presented
here paint yet a different picture. While (for reasons discussed
in the text) it is difficult to rule out the existence of a
continuous divergent contribution, we argue that the strong
divergence in $\rho(E\rightarrow 0)$ seen in previous numerical
work is actually indicative of a delta-function divergence at the
Fermi level.  In the final stages of writing this work, we became
aware of a recent conserving weak-localization
calculation\cite{yang} which  comes to the same conclusion.

The paper is organized as follows: in section \ref{sec:gfn} we
derive expressions for the Green's functions $\hat G(\r,\omega)$
needed to evaluate the two-impurity T-matrix.  Asymptotic
expressions for large $\r$ have been found
previously\cite{joynt,BalatskySalkola,pepinlee} and our results,
 valid for small $\omega$, are complementary. In section
\ref{sec:dos} we specialize to the fully nested strong scattering
model,  evaluate the density of states for different impurity
configurations, and show that there are three different classes of
impurity-pair orientation. For two of these classes, the DOS
diverges as $\omega \rightarrow 0$, while the DOS vanishes for the
third.  In all three cases, interference between impurities is
substantial as in Ref.\ \cite{pepinlee}, but ultimately the
observed divergences arise from the local rather than nonlocal
correlations.  In section \ref{sec:tau3} we establish a connection
between the zero-energy LDOS of the fully nested disordered system
and the zero-energy DOS of the one- and two-impurity problems. For
the fully nested model, we find that,  in a given configuration,
only impurities on a given sublattice contribute to the resonant
weight at zero energy. The impurities in this class form a network
with spatial separations equivalent to the resonant configurations
in the 2-impurity case, and numerical scaling of the total
spatially integrated DOS is shown to be consistent with
$\rho(\omega) \sim \delta(\omega)$.

In section \ref{sec:conc}, we summarize our conclusions and
discuss the less symmetric situation found in the cuprates.
 We argue that, because the STM
averages over many multi-impurity eigenstates, the LDOS indeed
appears to represent a set of nearly isolated impurity states with
spectral features which are similar from impurity to impurity. On
the other hand, we expect the width of these local states in
energy to be typically the impurity bandwidth arising from the
full disordered system.

\section{Two impurities in a half-filled band}
\label{sec:2imp}
\subsection{Green's functions}
\label{sec:gfn}

The BCS Hamiltonian for a pure $d$-wave singlet superconductor in a
tight-binding band can be written as:
\begin{subequations}
\begin{eqnarray}
    H_{0}&=&\sum_{\k}\Phi_{{\k}}^{\dagger}
    [(\epsilon_{\k}-\mu) \hat \tau_{3}+\Delta_{k}\hat \tau_{1}]\Phi_{\k},
    \label{hama}\\
    \epsilon_\k&=&-2t(\cos k_x +\cos k_y),\\
    \label{hamb}
    \Delta_\k &=& \Delta_0(\cos k_x -\cos k_y),
    \label{hamc}
\end{eqnarray}
\end{subequations}
where $\Phi_{\k}=(c_{\k\downarrow} c^{\dagger}_{-\k\uparrow} )$ is a
Nambu spinor, and $\hat \tau_{i}$ are the Pauli matrices.  Energies
are measured relative to the center of the band, so a chemical
potential of $\mu = 0$ corresponds to half-filling.  The associated
Green's function is, in real space, a function of the relative
coordinate ${\bf r} = (m,n)$, where $\rr$ is measured in units of the
lattice constant and $m$ and $n$ are integers:
\begin{eqnarray}
\hat G^0(\r,\omega) &=& \sum_\k e^{i\k\cdot \r} \hat G^0(\k,\omega),
\nonumber  \\
&=& \sum_{\k} \cos(k_x m)\cos(k_y n)  \nonumber \\
&&\times
\frac{\omega \hat \tau_0 + (\epsilon_k-\mu)\hat \tau_3 + \Delta_\k \hat \tau_1}
{\omega^2 - E_\k^2} ,
\label{gr1}
\end{eqnarray}
where $E_\k = \sqrt{\epsilon_\k^2 + \Delta_\k^2}$ denote quasiparticle
energies, $\hat {~}$ denotes a matrix in Nambu space and the
superscript denotes the bare Green's function.  Frequently, it is
convenient to make the decomposition in terms of Nambu spinors
\[
\hat G^0(\r,\omega) = \sum_{j = 0}^3 G^0_j(\r,\omega) \hat \tau_j.
\]
An ensemble of $N$ short-range scattering potentials at a set of sites
$\R_i$ introduce a perturbation
\begin{eqnarray*}
    H_{imp}&=& V_0 \sum_{i=1}^N
    \Phi_{\R_i}^{\dagger}\hat \tau_3\Phi_{\R_i}
\end{eqnarray*}
where $V_0$ is the strength of the impurity potential.
Formally, there is an exact solution for the disordered Green function in
terms of the $2N \times 2N$ many-impurity T-matrix:
\begin{eqnarray*}
\hat G(\r,\r^\prime,\omega)&=& \hat G^0(\r - \r^\prime,\omega) \\
&& + \sum_{i,j}
\hat G^0(\r -\R_i,\omega) \hat T_{ij}(\omega) \hat G^0(\R_j - \r,\omega)
\end{eqnarray*}
with $i,j$ the position indices of the impurity sites, and
\[
{\bf \hat T} = [ {\bf 1} \otimes \hat \tau_0 - \hat \tau_3 V_0 {\bf \hat G^0}
     (\omega)]^{-1} {\bf 1} \otimes \hat \tau_3 V_0,
\]
where the boldface indicates a matrix in spatial indices in the
subspace of impurity sites (ie.\ ${\bf \hat G^0}_{ij}(\omega) = \hat
G^0(\R_i-\R_j,\omega)$) and the inverse is a matrix inverse.  In the
limit of a single impurity, the T-matrix simplifies to $\hat T(\omega)
= [ V_0^{-1} \hat \tau_3 - \hat G^0(0,\omega) ]^{-1}$, with $\hat
G^0(0,\omega) \equiv \hat G^0(\r=0,\omega)$. This limit has been
studied extensively.

In this work, we are particularly interested in the two-impurity
T-matrix with one impurity at the origin (for simplicity) and the
other a displacement $\R = (m,n)$ from the origin.  The 2-impurity
T-matrix is a $4\times 4$ matrix which satisfies
\begin{equation}
{\bf \hat T} = \left [ \begin{array}{cc}
  V_0^{-1}\hat \tau_3 - \hat G^0(0,\omega) & -\hat G^0(\R,\omega) \\
  -\hat G^0(\R,\omega) & V_0^{-1} \hat \tau_3 - \hat G^0(0,\omega)
\end{array} \right ]^{-1}
\label{generaltmat}
\end{equation}
Expressions for the local Green's function $\hat G^0(0,\omega)$ have been
derived in many places, but the nonlocal Green's function
$\hat G^0(\R,\omega)$ is less well understood, although several
asymptotic expressions have been
found\cite{joynt,BalatskySalkola,pepinlee}.  In the Appendix, we
derive expressions which are valid for the half-filled band, and
which become exact in the limit $\omega \rightarrow 0$.

We find that the local Green's function for general complex
$\omega$ is
\begin{equation}
\hat G^0(0,\omega) = -\frac{\alpha\omega}{2} \ln \frac{\Lambda^2}{-\omega^2}
\hat \tau_0,
\label{eqG0}
\end{equation}
where $\alpha = N/(2\pi v_F
v_\Delta)$, $N=4$ is the number of nodes, $v_F$ is the Fermi
velocity and $v_\Delta$ is the anomalous quasiparticle velocity
$|\nabla_\k \Delta_\k|$, and the cutoff $\Lambda$ is of order
$\Delta_0$.
The expansion in $\omega$ for
$\r=(m,n)$ depends on whether $n$ and $m$ are odd or even.
For the (even,even) case, we have
\begin{equation}
\hat G^0(\r,\omega) \rightarrow (-1)^{\frac{n+m}{2}} \left [
G^0_0(0,\omega) + \omega C_0(\r)
\right
] \hat \tau_0, \label{eveneven}
\end{equation}
where $C_0(\r)$ is a real function of $\r$.
We find similar leading-order expressions for $(m,n) = \mathrm{(odd,odd)}$,
\begin{equation}
\hat G^0(m,n,\omega) \rightarrow \omega C_0(\r) \hat \tau_0,
\label{oddodd}
\end{equation}
while for $(m,n) = \mathrm{(odd,even)}$ or (even,odd),
\begin{equation}
\hat G^0(m,n,\omega) \rightarrow C_1(\r) \hat \tau_1 + C_3(\r) \hat \tau_3,
\label{oddeven}
\end{equation}
where $C_1(\r)$, and $C_3(\r)$ are real constants.
This distinction between even and odd sites accounts for the
oscillatory nature of the wavefunctions for the special case that the
Fermi wavevector is commensurate with the lattice.

\subsection{Density of states for two impurities}
\label{sec:dos}
\begin{figure}[tb]
\begin{center}
\leavevmode
\includegraphics[width=.9\columnwidth]{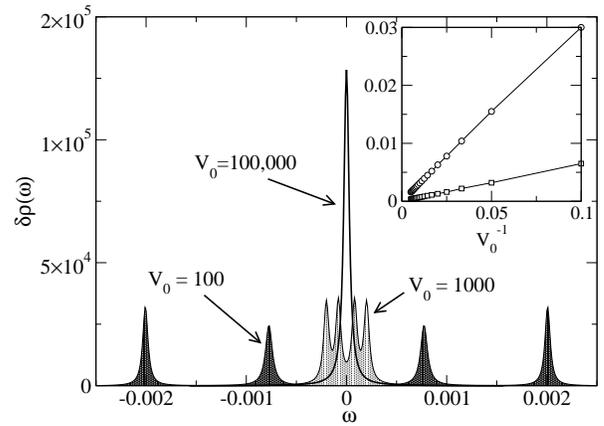}
\caption{Change in the quasiparticle density of states arising from
impurities separated by $\R=(2,2)$ as a function of energy $\omega/t$
for scattering potential $V_0=10^6t$. {\em Inset:} Scaling of the
resonance peak energies as a function of $V_0$. The DOS for $\R=(2,0)$
is almost identical. Energies are measured in units of $t$, and
$\Delta_0 = 0.1t$. }
\label{figR22}
\end{center}
\end{figure}

In this section, we derive expressions for the density of states
for two impurities in a half-filled band.  The discussion focusses
on the unitary limit $V_0 \rightarrow \pm \infty$.  The
half-filled tight-binding band possesses a particular global
nesting symmetry \cite{yashenkin}
$\hat \tau_2 \hat G^0(\k+\Q,\omega)\hat \tau_2= -\hat G^0(\k,\omega)$,
with $\Q=(\pi,\pi)$ which is satisfied at $\omega = 0$.  For
simplicity, we call this the $\tau_2$ symmetry.
Potential scattering violates this symmetry, but in the case of
infinite potential, impurity sites are effectively removed from the
lattice, and the symmetry is recovered for any disorder configuration
at $\omega=0$.
In real space (see eg.\ \cite{otherdivergence})the $\tau_2$ symmetry
may be expressed as
\begin{equation}
\hat \tau_2 \hat G(\r,\r^\prime,\omega)\hat \tau_2
= -e^{i\Q\cdot (\r-\r^\prime)}
\hat G(\r,\r^\prime,\omega).
\label{eqtau3}
\end{equation}
It will be useful to decompose the square tight-binding lattice
into the usual two interleaved sublattices (denoted A and B). The
phase factor on the right hand side of Eq.\ (\ref{eqtau3}) is $+1$
if $\r$ and $\r^\prime$ belong to the same sublattice, and $-1$
otherwise.

The simplest quantity of interest is the quasiparticle density of
states
\begin{eqnarray*}
    \rho(\omega) &=& \sum_n [\delta(\omega-E_n) +
    \delta(\omega+E_n)] \\
    &=& \rho_0(\omega) + \delta \rho(\omega),
\end{eqnarray*}
where $E_n$ are the positive energy eigenvalues of the superconducting
Hamiltonian, $\rho_0(\omega)$ is the DOS of the disorder-free
system and $\delta \rho(\omega)$ is the change induced by the
impurities.  The DOS is related to the 2-impurity T-matrix defined in
Eq.\ (\ref{generaltmat}) by the phase shift $\eta(\omega)$\cite{hewson}:
\begin{equation}
    \delta \rho(\omega) =
    \frac 1\pi \frac{\partial \eta}{\partial \omega},
    \label{eq:DOS}
\end{equation}
where $\eta$ is given by,
\begin{equation}
    \eta(\omega) = \tan^{-1}\frac{\mbox{Im }
    \det {\bf \hat T}}{\mbox{Re } \det {\bf \hat T}}
    \label{eq:phase}
\end{equation}
and the determinant is over spatial and spin indices.

We start with a discussion of two impurities
belonging to one of the sublattices.  The two impurities are at $\R_1$ and
$\R_2$ with $\R \equiv \R_1 - \R_2 = (m,n)=$(even,even) or (odd,odd).
The two-impurity T-matrix defined in Eq.\ (\ref{generaltmat}) is
particularly simple in this case:
\[
{\bf \hat T} = \frac 1 D \left[ \begin{array}{cc}
    -G^0_0(0,\omega) \hat \tau_0 & G^0_0(\R,\omega) \hat \tau_0 \\
    G^0_0(\R,\omega)\hat \tau_0 & -G^0_0(0,\omega) \hat \tau_0
    \end{array} \right]
\]
where $D = G^0_0(0,\omega)^{2}-G^0_0(\R,\omega)^2$.
Noting that
\[
    \det {\bf \hat T} = \frac{1}{D^2},
\]
we keep the leading order terms in $G_0(\R,\omega)$ as
$\omega \rightarrow 0$, given explicitly in Eq.\ (\ref{eveneven}) and
(\ref{oddodd}) and find that
$\det {\bf \hat T}$ diverges as
\[
    \det {\bf \hat T} \rightarrow \left \{
    \begin{array}{ll}
    [2\omega C_0(\R)G^0_0(0,\omega)]^{-2} & \R = (\mathrm{even, even}) \\
    G^0_0(0,\omega)^{-4} & \R = (\mathrm{odd, odd})
    \end{array}
    \right .
\]
and (analytically continuing $\omega$ to the real axis)
\begin{equation}
    \delta \rho(\omega)\rightarrow \left \{
    \begin{array}{ll}
    1/[\omega \log^2(\Lambda/\omega)] & \R = (\mathrm{even, even}) \\
    2/[\omega \log^2(\Lambda/\omega)] & \R = (\mathrm{odd, odd})
    \end{array}
    \right .
\label{divergentdos}
\end{equation}
Because of the similarity of the approaches, we are able to compare
our findings with those of PL\cite{pepinlee} in some detail.
Although the form of Eq.\ (\ref{divergentdos}) is suggestive of the
asymptotic result of PL for the disorder-averaged
limit, its origin is quite different.  This difference is easiest
to see for the (odd,odd) impurity configuration: here the local
Green's function $\hat G^0(0,\omega)$ is dominant over the
nonlocal term $\hat G^0(\R,\omega)$ and the physics of the low
energy resonance is essentially that of two non-interacting
impurities.  The total weight of the resonance is therefore twice
that of a single impurity.  For the (even,even) case the situation
is a little more complicated, since the local and nonlocal terms
are nearly equal in magnitude; interference effects reduce the
spectral weight of the combined resonance to half that of two
isolated resonances. In both cases the situation is quite
different from Ref.\ \cite{pepinlee} where the logarithmic
divergence arises from averaging over all possible impurity
separations using the approximate form $\hat G^0(\R,\omega) \sim
1/R$ out to a cutoff $\sim t/R$.  The PL result is inherently
nonlocal.

Numerical calculations for two impurities with separation
$\R=(2,2)$ are shown in Fig.\ \ref{figR22}.  For $V_0 = 100t$,
four clearly defined peaks are seen, corresponding to the level
splitting of the single impurity resonances of the isolated
impurities\cite{zhutwoimpurity}.  As shown in the inset, the peak
positions scale strongly with $V_0$, and a single peak appears
only when $V_0 \sim 10^5t$.

\begin{figure}[tb]
\begin{center}
\leavevmode
\includegraphics[width=.9\columnwidth]{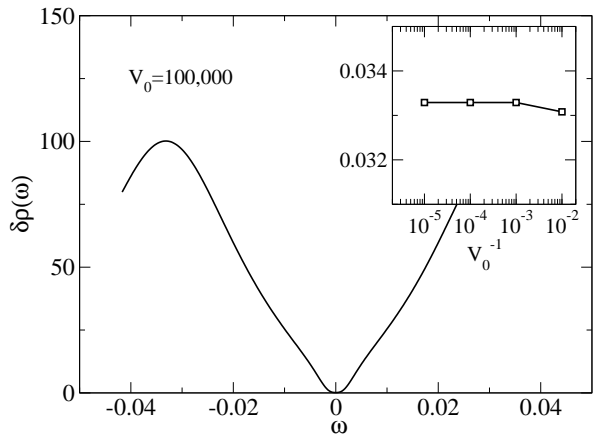}
\caption{Change in the quasiparticle density of states for
impurities separated by $\R=(2,1)$ as a function of  energy
$\omega/t$ for scattering potential $V_0=10^6t$ and
$\Delta_0=0.1t$. {\em Inset:} Scaling of the resonance peak
position as a function of $V_0$.} \label{figR21}
\end{center}
\end{figure}

We continue now  with the case where the impurities
belong to different sublattices and are separated
by $\R = $(even,odd).  The two-impurity T-matrix defined in Eq.\
(\ref{generaltmat}) is:
\[
{\bf \hat T} = \frac{1}{D^\prime} \left[ \begin{array}{cc}
    -G^0_0(0,\omega) \hat \tau_0 & C_1(\R) \hat \tau_1+ C_3(\R) \hat \tau_3\\
    C_1(\R) \hat \tau_1+ C_3(\R) \hat \tau_3 & -G^0_0(0,\omega) \hat \tau_0
    \end{array} \right]
\]
with $D^\prime = G(0,\omega)^2 - C_1(\R)^2 - C_3(\R)^2$.
It follows easily that $\det {\bf \hat T} = {D^\prime}^{-2}$ and that
\begin{equation}
\delta \rho(\omega\rightarrow 0) \propto \frac{d}{d\omega}\left (
\omega^2 \ln \frac{\Lambda}{\omega} \right ) \rightarrow 0
\end{equation}
A similar result holds for $\R =$(odd,even).  Physically, the fact
that $\delta \rho$ vanishes at the Fermi level indicates that bound
state energies must always arise at nonzero energies.  Numerical
calculations of the DOS shown in Fig.\ \ref{figR21} demonstrate that
there is no remnant of the single impurity $\omega \rightarrow 0$
divergence for this orientation, and that the resonance energies scale
very little with $V_0$.  In this case, it is the dominance of the
nonlocal terms which shifts the resonance to finite energy.

\section{Disordered system with global particle-hole symmetry}
\label{sec:tau3}

In this section, we discuss the correspondence between the two
impurity problem and the disordered $d$-wave superconductor.  There
are two separate issues to be dealt with.  The first has to do with
the nature of the divergence at $\omega = 0$ which occurs in the
tight-binding model, while the second has to do with the more general
question of how the impurity band evolves with impurity concentration.
For these calculations, we numerically diagonalize the mean-field
Hamiltonian for a random distribution of impurities,  under the
assumption of a homogeneous order parameter for a finite sized $L
\times L$ system with periodic
boundaries. For a detailed description of the method, we refer the
reader to eg.\ \cite{atkinsonops}.  We retain the eigenenergies $E_n$
and the eigenvectors
\[
\Psi^{(n)}(\r) = \left ( \begin{array}{c} u^{(n)}(\r) \\[2mm]
v^{(n)}(\r)
  \end{array} \right ).
\]
The total density of states is just $\rho(\omega) = \sum_n \delta(\omega-E_n)$,
and the single-spin LDOS is
\[
\rho({\bf r},\omega) = \sum_n |u^{(n)}({\bf r})|^2
\delta(\omega-E_n).
\]
Since there is no moment formation, $\sigma=\uparrow$ and
$\sigma=\downarrow$ are equivalent.

The Green's function $G^0(\k,\omega=0)$ for the Hamiltonian
(\ref{hama}) (with $\mu = 0$) has the special symmetry $\hat \tau_2
\hat G^0(\k+\Q,0) \hat \tau_2 = -G^0(\k,0)$ where $\Q=(\pi,\pi)$ is
the antiferromagnetic wavevector.  The $\tau_2$ symmetry is
required\cite{atkinsondetails,yashenkin} for the divergence in
$\rho(\omega\rightarrow 0)$.  This symmetry is only strictly satisfied
when $L$ is even\cite{parity},
and an even-odd oscillation at the Fermi
level as a function of $L$ is clearly evident in our numerical work.
Throughout this paper, we restrict ourselves to even $L$.

\subsection{Divergence at $\omega = 0$}

\begin{figure}[floatfix]
\begin{center}
\includegraphics[width=\columnwidth]{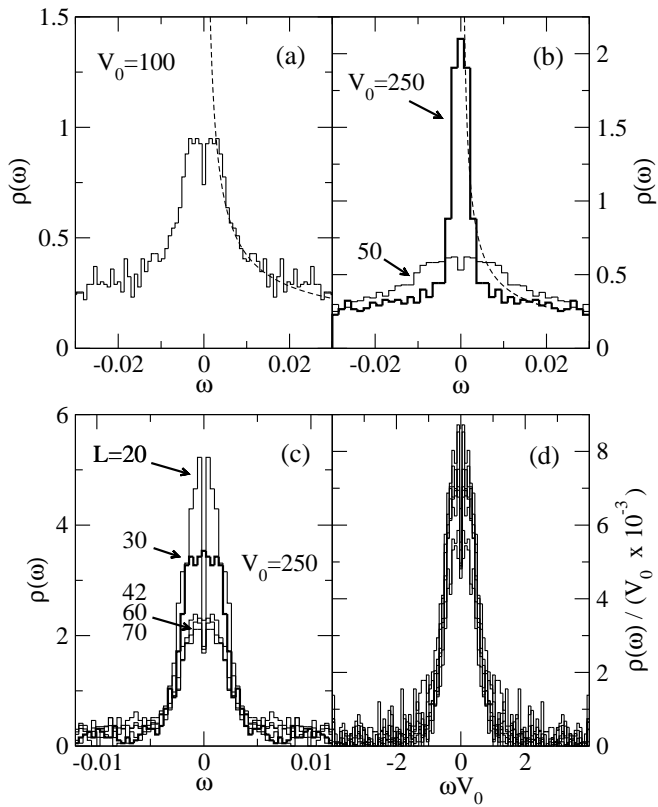}
\caption{Total density of states for $n_i = 0.1$.
(a) DOS for $V_0 =
100t$ (solid) and $L=60$.  Eq.\ (\protect \ref{eqPL}) is plotted for comparison
(dashed line).
(b) Scaling of DOS with $V_0$.  $\rho_{PL}(\omega)$ is again plotted
for comparison. (c) Scaling
of DOS with $L$.  (d) Scaling of DOS with $V_0$ for $V_0/t = 100,\,
500,\, 1000,\, 5000,\, 10^4,\,10^5, \, 10^6$ and $L=60$.  A background
$\rho_0 = 0.25 t^{-1}$ has been subtracted.  The figure shows that the
density of states is a peaked function whose width scales as $1/V_0$
and whose height scales as $V_0$, suggesting that
$\lim_{V_0\rightarrow \infty} \rho(\omega) \sim \delta(\omega)$.
All energies are in units of $t$.}
\label{figtotdos}
\end{center}
\end{figure}

The DOS for a large concentration $n_i = 0.1$ of strong scattering
impurities in a $d$-wave superconductor is shown in Fig.\
\ref{figtotdos}.  The figure is restricted to low energies, and shows
only the zero-energy peak at the Fermi level, and a small portion of
the impurity band.  For comparsion, the $d$-wave gap has an energy
$\Delta_0 = 0.2t$ and the gap edge in the tunneling density of states
is $0.4t$.  For clarity, we often make a distinction between states in
the peak and states in the impurity band, by which we mean states
belonging to the DOS plateau which is characterized by a constant
density of states $\rho_0$.  In Fig.\ \ref{figtotdos}(a), for example,
$\rho_0 \approx 0.25 t^{-1}$.

In Fig. \ref{figtotdos}(a), the total DOS is shown for an impurity
potential $V_0 = 100t$ corresponding to a strong scattering
potential. The results are in quantitative agreement with earlier
numerical work\cite{atkinsondetails,Zhu}.  The PL result
\begin{equation}
\rho_{PL}(\omega) \approx \frac{n_i }{|\omega| [\ln^2(\Lambda/\omega)
+ (\pi/2)^2]},
\label{eqPL}
\end{equation}
is also shown.  Here, we take $\Lambda = 1$, first because this was
the cutoff used in previous numerical work\cite{Zhu} and, second,
because this gives a good fit to the numerics at $V_0=100t$.  It
should be clear from Figs.\ \ref{figtotdos}(a) and (b) however, that although
the fit is striking at $V_0=100t$, it is less so for other values of
$V_0$.  In our numerics, we find a smooth evolution of the low energy
peak as a function of $V_0$ and there is no value of $V_0$ beyond which
the asymptotic behaviour saturates. In general, $\rho_{PL}(\omega)$ does
not appear to fit the data well, except for certain special parameter sets.
\begin{figure}[floatfix]
\begin{center}
\includegraphics[width=\columnwidth]{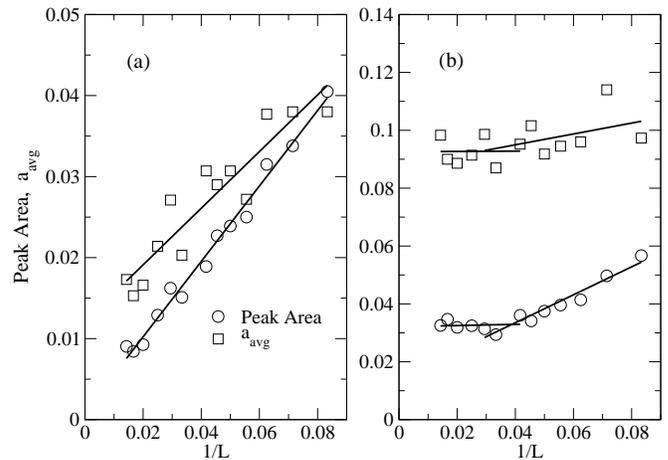}
\caption{Scaling of the peak area and inverse participation ratio
with system size $L$ and $V_0 = 10^6t$.  $a_{avg}$ is averaged
over states with energies $10^{-5}t < E_n < 0.03t$.  (a) Scaling
for $n_i=0.1$. (b) Scaling for $n_i = 0.2$. Solid lines are linear
fits to the data. For these curves, $\Delta_0 = 0.5t$. }
\label{figscaling}
\end{center}
\end{figure}
The shape of the peak at $\omega=0$ is modified by finite-size
effects.  There is a crossover in behaviour which occurs when the mean
level spacing in the impurity band $\delta_L = 1/(\rho_0 L^2)$ is
comparable to the peak width.  Scaling of the DOS is shown for
$V_0=250t$ in Fig.\ \ref{figtotdos}(c).  The peak height scales with
$L$ for $L \lesssim 40$ and saturates at larger system sizes.  The
implication is that some care must be taken in approaching the $V_0
\rightarrow \infty$ limit.

The unitary limit of the infinite system may be approached in two
ways.  First, one may consider taking $\lim_{V_0\rightarrow
\infty} \lim_{L\rightarrow \infty}$
so that the level spacing in the impurity
band is much less than the peak width.  Second, one may consider
taking the limit $L\rightarrow \infty$ with $V_0=\infty$.  In the
first approach, the $\tau_2$ symmetry is only strictly satisfied
when $L=\infty$, while in the second approach, the
$\tau_2$ symmetry is rigorously satisfied for any even value of
$L$.  For this reason, we view the second approach as preferable.

The limit $V_0\rightarrow \infty$ for fixed $L$ is illustrated in
Fig.\ \ref{figtotdos}(d).  The data are scaled by the impurity
potential, and the general trend is that as $V_0$ is increased, a
sharp peak develops at $\omega = 0$.  Furthermore, the peak scales as
$\rho(\omega) \approx V_0 F(\omega V_0)$, implying that
\begin{equation}
\lim_{V_0\rightarrow \infty} \rho(\omega) \sim
\delta(\omega).
\label{deltapk}
\end{equation}

Not surprisingly, the weight contained in the delta-peak in the $V_0
\rightarrow \infty$ limit scales with $L$, as shown in Fig.\
\ref{figscaling}.  For $n_i=0.1$, this scaling is consistent with what
we found in Fig.\ \ref{figtotdos}(c).  When $n_i=0.2$, on the other
hand, the peak area saturates when $L \gtrsim 40$, which is not
expected since the peak width is still many orders of magnitude
smaller than the typical level spacing $\delta_L$ in the impurity
band.  To learn more about the origin of this saturation we plot in
the same figure the scaling of the inverse participation ratio,
defined by
\[
a(\omega) = \sum_n \frac{\sum_{i} [ u^{(n)}(\rr_i)^4 +
v^{(n)}(\rr_i)^4] } {\big (\sum_i [u^{(n)}(\rr_i)^2 +
v^{(n)}(\rr_i)^2] \big )^2}\delta(\omega-E_n).
\]
$a(\omega)$ scales as $L^{-d}$ for wavefunctions which are extended in
$d$-dimensions, and does not scale with $L$ for localised states.
The localization length is typically extracted from the
crossover which occurs when
$L\approx \xi_L$ where $\xi_L$ is the localization length.  As we
shall see below, states in the delta-peak behave differently from
those in the impurity band, and we find that the peak area is
correlated with the localization properties of the impurity band.  In
Fig.\ \ref{figscaling}, the inverse participation ratio is averaged
over states in a narrow energy window adjacent to (but not including)
the delta peak. It is evident from the figure that for $n_i = 0.2$, a
crossover to the localized regime occurs, and we can extract a
localization length $\xi_L \approx 40$.
\begin{figure}[floatfix]
\begin{center}
\includegraphics[width=\columnwidth]{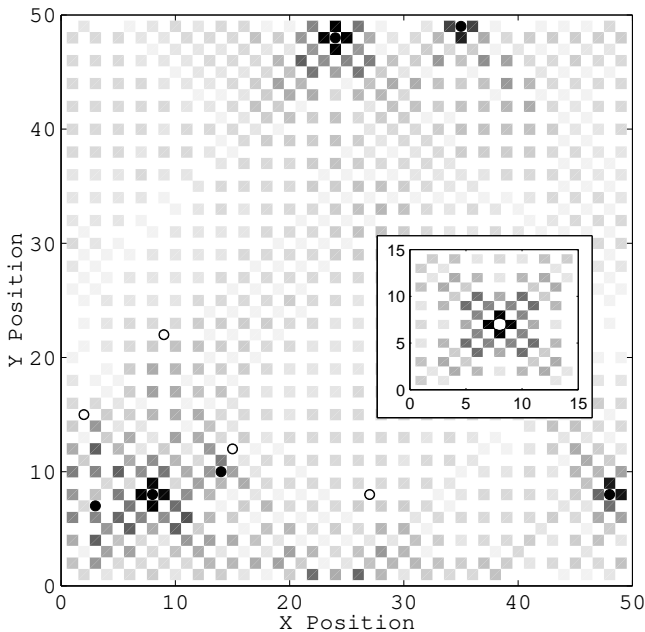}
\caption{Local density of states for 0.4\% concentration of impurities
and $|E_n| < 10^{-5}t$ (4 eigenvalues).  Impurity locations on
sublattice A are indicated with open circles, those on sublattice B
with filled circles, and the impurity potential is $V_0 = 10^6t$. {\em
Inset:} The inset shows a detail of the LDOS for a single impurity.  }

\label{figldos4}
\end{center}
\end{figure}
Remarkably, we find that the area of the $\delta$-peak appears to
saturate when $L>\xi_L$.  This situation is analogous to one reported
earlier in $d$-wave superconductors possessing no special symmetries.
There, it was shown that quantum interference
(arising from ``maximally crossed'' diagrams)
leads to a suppression
of the DOS at the Fermi level\cite{fisher} over an energy scale
$\delta_{\xi_L} = 1/(\rho_0\xi_L^2)$ (This situation is illustrated in
Fig.\ref{figschematic}(b)).  In finite size systems, the energy scale for
the DOS suppression is actually $\delta_L$ and the scaling of the
suppression saturates when $L > \xi_L$\cite{atkinsondetails}.

\begin{figure}[floatfix]
\begin{center}
\includegraphics[width=\columnwidth]{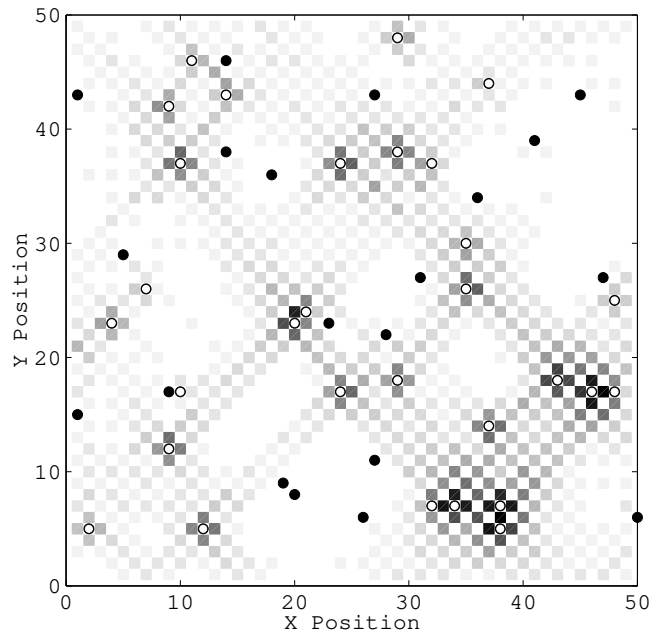}
\caption{Local density of states for 2\% concentration of impurities
and $|E_n| < 10^{-5}t$ (20 eigenvalues). Impurity locations on
sublattice A are indicated with open circles, those on sublattice B
with filled circles, and the impurity potential is $V_0 = 10^6t$.  }
\label{figldos2}
\end{center}
\end{figure}

We note that the origins of the delta-peak divergence are
fundamentally different from those discussed in PL, where the
divergence arises from the cumulative effects of interference between
a large number of distant impurities.  Here the result appears to be a
mesoscopic effect which survives because localization makes the
effective system size finite.  However, although the delta-peak result
is different from earlier predictions for a continuous divergence at
the Fermi level, it does not preclude the existence of an additional
divergent term which is unobservable because of finite system size
effects.  Indeed, if we consider the effect of finite system size on
the PL result we find the interference between distant impurities is
cut off by $L$ and we should make the substitution $\omega \rightarrow
\max(\omega,t/L)$ in $\rho_{PL}$, implying a cutoff energy $\omega_c
\approx t/L$ below which the DOS saturates.  By this estimate, the
contribution to the plot in Fig.\ \ref{figtotdos} is cut off below
$\omega_c \approx 0.017t$, suggesting that the PL peak should be
unobservable.

It is particularly instructive to consider the structure of the
delta-peak divergence in real space.  Figure \ref{figldos4} shows
the combined local density of states from the eigenstates with
energy $|E_n| < 10^{-5} t$ which make up the delta-peak (these
states are well-separated from all other eigenvalues).
 For a
single impurity (shown in the inset) the zero-energy resonance has
a fourfold spatial structure with bright lobes on sites adjacent
to the impurity along the antinodal (100) and (010) crystal
directions, and extended tails in the nodal (110) and
($1\overline{1}0$) directions, in agreement with many earlier
calculations\cite{Balatsky}.  For 0.4\% disorder (10 impurities),
the situation is quite different; even at this relatively low
concentration, there is significant interference between
impurities.  We see four pronounced zero-energy resonances, but
the remaining six impurities are---at best---only weakly visible.
For each of the visible resonances, the LDOS has the superficial
structure of the isolated impurity LDOS, with maxima appearing in
the antinodal direction
and tails extending away from the impurities in the nodal directions.
However, there is no obvious correlation between
the degree of isolation and the appearance of a zero-energy resonance.
Indeed, of the four strong resonances, only two are more than 10
lattice sites from the nearest impurity.  For 2\% disorder (50
impurities), shown in Fig.\ \ref{figldos2}, the situation is similar.
Only a small fraction of impurities contribute to the zero-energy LDOS
and, again, the visible resonances do not necessarily belong to the
most isolated impurities.  At this higher impurity concentration,
however, a definite pattern in the LDOS is observable.  Long tails
along the $(110)$ and $(1\overline{1}0)$ directions give the
appearance of a network of impurities.

\begin{figure}[floatfix]
\begin{center}
\includegraphics[width=\columnwidth]{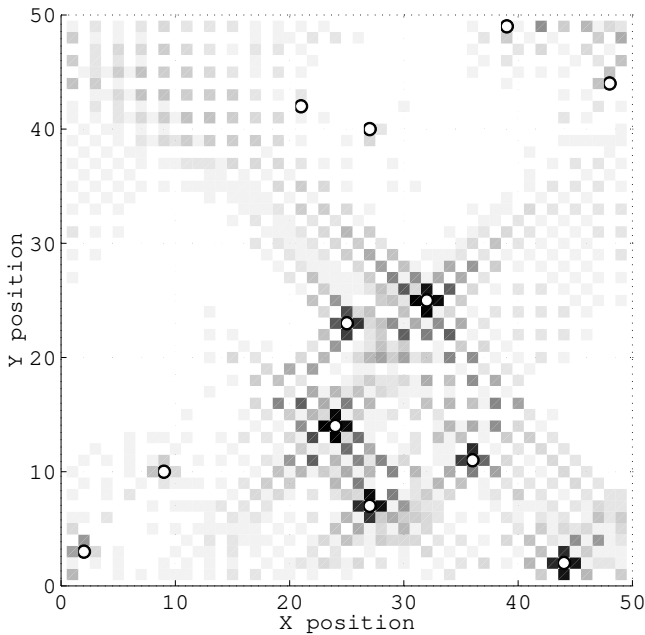}
\caption{Local density of states for 0.5\% impurities and $V_0 =
10^6t$ derived from a single eigenvalue with $E_n = 0.0318t$.}
\label{figldos3a}
\end{center}
\end{figure}

Remarkably, we find that all impurities within the visible network
in Fig.\ \ref{figldos2} belong to one sublattice, arbitrarily
denoted A, while the remaining impurities belong to the B
sublattice.
Similarly, in Fig. \ref{figldos4}, all visible impurities belong to
the B sublattice.
While this is reminiscent of the two-impurity problem
discussed in the previous section, it is also quite surprising.
For the two-impurity problem, it was shown that the zero-energy
resonance is preserved when both impurities inhabit the same
sublattice and is destroyed otherwise. The natural extrapolation
is that, for a random distribution of many impurities, every
impurity is expected to have have some reasonably close neighbor
belonging to the other sublattice which contributes to the
destruction of the zero-energy peak.  Clearly, this does not
happen.  Instead, the impurities belonging to the A sublattice for
this sample are dominant at $\omega=0$ for reasons we do not
completely understand at present.  An apparent consequence of this
dominance is that the resonances of impurities belonging to the B
sublattice are shifted to higher energies.
We speculate, but cannot prove, that the system in the thermodynamic
limit will have ``domains"
of typical size $\xi_L$
in which either A or B impurities are resonant.

The observed networks are also reminiscent of an earlier
proposal\cite{BalatskySalkola} that impurities form networks from
single impurity resonances which lead to a delocalization
transition as $\omega \rightarrow 0$. Numerical scaling
calculations\cite{Zhu} for a finite impurity potential ($V_0 =
100t$) did not find such a transition, however, nor does the
present work (see below).  In any case, we emphasize  that the
sharply defined networks exhibited above are a feature of
Hamiltonians with $\tau_2$ symmetry only, and not a general
feature of $d$-wave superconductors as suggested in
\cite{BalatskySalkola}.

\subsection{Impurity band away from $\omega = 0$}

\begin{figure}[floatfix]
\begin{center}
\includegraphics[width=\columnwidth]{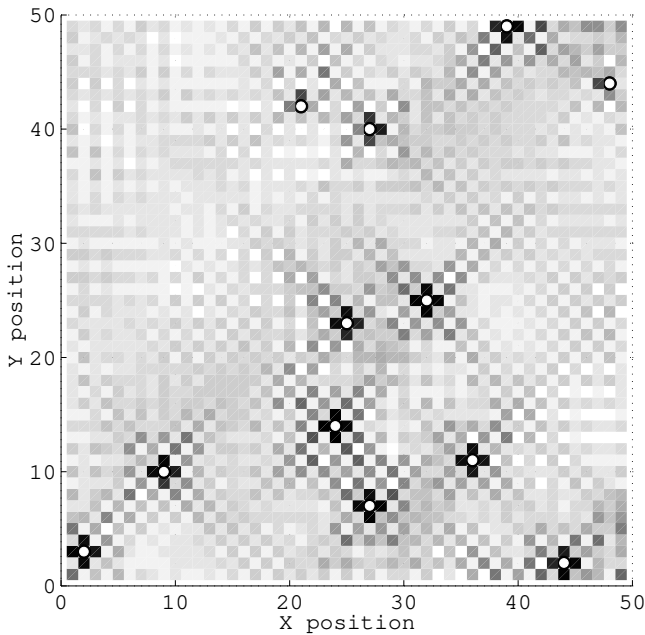}
\caption{Local density of states for 0.5\% impurities and
$V_0=10^6t$ averaged over 5 eigenvalues in the energy interval
$|E_n - 0.03t| < 0.02t$.  } \label{figldos3b}
\end{center}
\end{figure}

We now turn our attention to the states in the impurity band away
from $\omega = 0$.  The previous analysis raises some interesting
questions about the  formation of this ``band".  Is the $\omega>0$
DOS plateau formed, as Figures \ref{figldos4} and \ref{figldos2}
perhaps suggest, by summing over many impurities, some of which
are resonant at a given energy and others not?  This would imply
that, as energy was scanned in STM experiments, different
impurities would ``light up" -- become resonant--and turn off at
different energies within the impurity band, a scenario we will
refer to as ``inhomogeneous broadening" of the impurity
resonances.    Experimental data \cite{davisnative,davisZn}
indicate instead that all impurities, regardless of local
environment, appear to be resonant all through the impurity band,
i.e. that each local spectral function is qualitatively similar in
position and  width, i.e. ``homogeneous broadening". In addition,
there is some evidence from explicit Zn substitution\cite{davisZn}
that the number of impurity resonances corresponds closely to the
number of Zn atoms introduced into the crystal, i.e. there are no
atoms which do not light up.  It is for this reason that
interpretations have typically been given in terms of one-impurity
models.   However, in the same experiments the width of spectral
features is roughly an order of magnitude larger  than those
predicted  by the simplest one-impurity models.

These apparent paradoxes can be resolved by recognizing that the
energy range probed by STM, although very small
($\cal{O}$(0.1meV)) in laboratory terms, is still large enough to
sample an essentially infinite number of eigenstates of the
macroscopic system. In Fig. \ref{figldos3a} we show the LDOS
derived from a single eigenstate at an energy which is in the
impurity band, but away from the zero-energy delta-peak. Two
features of this figure stand out. First, as was the case at
$\omega=0$, only a fraction of the impurities contribute to any
given eigenstate. Second, the extended tails which were are
important in the formation of the delta-peak are blurred by the
incommensurability between the lattice and the wavevectors
contained in the eigenstate.  As we move further away from $\omega
=0$, this incommensurability becomes more pronounced and the tails
become increasingly blurred.

The inequivalency between impurities in Fig.\ \ref{figldos3a} is
surprising not only because STM provides little evidence for such a
picture, but also because the arguments about the formation of
networks fail when $\omega \neq 0$ (indeed, there is no visible
network in the figure). When one now averages the LDOS over a small
energy window, as in Fig.\ \ref{figldos3b}, the system starts to look
much more homogeneous, in the sense that all impurities contribute
visible resonances with the classic fourfold symmetry.  The window
width is small compared to the impurity band, and we have checked that
the pattern averaged in this way remains roughly the same up to
energies of order the impurity band itself, $\gamma\simeq 0.25t$ for
the parameter set of the figure. Thus, it appears as if there is an
important distinction between individual eigenstates which determine,
for example, localization properties, and averages over finite energy
windows which determine the tunneling spectrum.

\begin{figure}[floatfix]
\begin{center}
\includegraphics[width=\columnwidth]{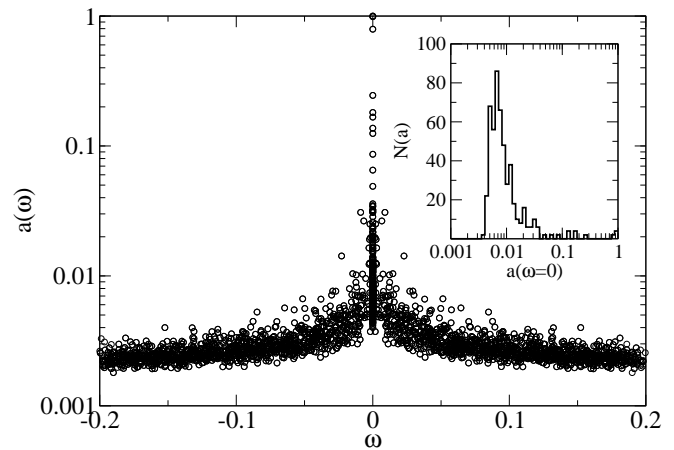}
\caption{Inverse participation ratio $a(\omega)$ for $n_i = 0.1$,
$V_0 = 10^6$ on a $30\times 30$ lattice with 50 configurations.
States in the impurity band for $|E_n| \gtrsim t/V_0$ have
approximately uniform spatial extent. States with $|E_n|V_0
\lesssim t$ exhibit strong fluctuations in spatial extent.  Note
that states with $a(E_n) = 1$ are confined to a single site, while
states with $a(E_n) \sim N^{-1}$ have a uniform spatial
distribution.  {\em Inset:} A histogram of the distribution of
$a(\omega)$ for states in the zero-energy $\delta$-function. Note
the logarithmic horizontal axis.} \label{figinvpart}
\end{center}
\end{figure}

Finally,
in order to solidify the connection between local and bulk
properties of the disordered system, the energy dependence of the
inverse participation ratio is plotted in Fig.\ \ref{figinvpart}.
For each impurity configuration, $a(E_n)$ is calculated for all
the eigenstates in the spectrum, and the aggregate is shown for 50
impurity configurations in the figure.  There is a clear
distinction between states inside and outside of the
$\delta$-peak.  States outside the $\delta$-peak are clearly
extended (the localization length is much larger than the system
size) and the distribution of $a(\omega)$ is relatively narrow at
a given energy.  On the other hand, there is a broad distribution
of $a(\omega)$ in the $\delta$-peak, indicating a mix of localized
and extended states.  The figure inset shows a histogram of the
distribution that demonstrates that most of the spectral weight in
the $\delta$-peak comes from the extended tails of the
resonances (Fig. \ref{figldos2}) and not from the highly-visible
localized resonances.

\section{Conclusions}
\label{sec:conc}

In this work we have studied the unitary limit of a disordered,
 half-filled $d$-wave superconductor with tight binding band.  This model
has a particular symmetry which is known to lead to a divergence
in the density of states at the Fermi level, although the
particular form of the divergence is controversial.  We began with
a discussion of the two-impurity problem, which yields an
analytical solution in the $\omega\rightarrow 0$ limit.  We found
that, owing to the commensurability of the nodal wavevectors and
the tight-binding lattice, there is an even-odd oscillation in the
two-impurity density of states in the unitary (infinite scattering
potential) limit. Impurity pairs on the same sublattice  have a
zero-energy divergence in the DOS similar to the single-impurity
divergence.  The origin of this divergence is quite different from
that reported earlier\cite{pepinlee}, which arises from the
cumulative interference of a large number of distant impurities.

We also noted that for impurities located on different
sublattices, the zero-energy single-impurity resonance is shifted
to higher energies as a result of interference.  Based on this
result alone, it is natural to assume that, in the many impurity
limit, any remnant of the single impurity peak will be obliterated
since each impurity is expected to have at least one reasonably
near neighbor which lies on the other sublattice.  Surprisingly,
we found that this is not the case.  Exact numerical studies of
finite-size systems show that unitary impurities actually form two
interleaved networks on the A and B sublattices, one of which
contains spectral weight at $\omega=0$, while the other does not.
Away from $\omega=0$, quasiparticle eigenstates are no longer
commensurate with the lattice, networks connecting resonant states
along the nodal directions are smeared, and individual eigenstates
consist of distorted resonances, which are inhomogeneously distributed.
  When the LDOS is averaged over a small
window in energy, however, as in an STM experiment, the fourfold
nature of the 1-impurity resonances is qualitatively recovered,
and resonances on individual impurity sites appear remarkably
similar, provided the impurities are not in immediate proximity.
Although the resonance peak positions may be qualitatively related
to the resonant energies of the underlying 1-impurity model, the
widths are very different, of order the impurity bandwidth, given
in the unitarity limit by $\gamma \simeq n_i\sqrt{\Delta_0 E_F}$.

The $\omega>0$ states of the tight-binding band are generic in the
sense that they do not posses the $\tau_2$ symmetry, or any other
symmetry which is not present in high $T_c$ superconductors. In
this sense, our results should be qualitatively applicable to the
experiments on real cuprate materials.  They suggest that the
ability of one-impurity models of any kind  to explain the details
of local STM spectra in samples with per cent level disorder are
severely limited.   To substantiate this picture, it will be
useful to compare local spectra on sites (e.g. impurity or nearest
neighbor sites) around different impurities using realistic bands.
Numerical calculations to realize the large systems necessary to
obtain the resolution required to reach definite answers to these
questions are in progress.

 {\it Acknowledgements} This work was partially supported by NSF grant
NSF-DMR-9974396, the Alexander von Humboldt Foundation, and by
Research Corporation grant CC5543.  PH and LZ would also like to thank
J. Mannhart and Lehrstuhl Experimentalphysik VI of the University of
Augsburg for hospitality during preparation of the manuscript.

\section{Appendix}
The purpose of this appendix is to derive expressions for the Green's
function $G(\R,\omega)$ with $\R = (m,n)$, which are valid in the
$\omega\rightarrow 0$ limit.  The starting point is Eq.\ (\ref{gr1}),
and the first step is to express
\begin{eqnarray*}
\cos(k_x m) &=& 2^{m-1} \cos^m k_x \\
&&+ \frac 12 \sum_{j=1}^{[\frac{m}{2}]}
(-1)^j \frac{m(m-j-1)!}{j!(m-2j)!} (2 \cos k_x)^{m-2j}
\end{eqnarray*}
where $[\ldots]$ refers to the integer part of the argument.  We focus
on the half-filled case $\mu=0$ and write Eq.\ (\ref{gr1}) as the sum
of terms of the form
\begin{eqnarray}
g_{pq} &=& \sum_\k \cos^p(k_x)\cos^q(k_y)
\frac{\omega \hat \tau_0 + \epsilon_k \hat \tau_3 + \Delta_\k \hat \tau_1}
{\omega^2 - E_\k^2}
\nonumber
\end{eqnarray}
where $p = m,m-2,\ldots$ and $q = n,n-2,\ldots$.  We proceed by
linearizing the dispersion near
the node at $(\pi/2,\pi/2)$ and making the coordinate transformation
$E^2 = \epsilon_\k^2 + \Delta_\k^2$, $\tan \theta =
\Delta_\k/\epsilon_\k$.
\begin{eqnarray}
g_{pq} &=&
\frac{\alpha} {2^{p+q}}
\int_0^{2\pi} \frac{d\theta}{2\pi}
\left( -\frac{\sin\theta}{\Delta_0} - \frac{\cos\theta}{2t} \right )^q
\left( \frac{\sin\theta}{\Delta_0} - \frac{\cos\theta}{2t} \right )^p
\nonumber \\
&&\times
\int_0^\Lambda E^{p+q+1}  dE
\frac{\omega \hat \tau_0 + E(\cos\theta \hat\tau_3 + \sin \theta \hat\tau_1)}
{\omega^2 - E^2}, \nonumber
\label{gr2}
\end{eqnarray}
The prefactor is $\alpha = N/(2\pi v_F
v_\Delta)$ where $N=4$ is the number of nodes, $v_F$ is the Fermi
velocity and $v_\Delta$ is the anomalous quasiparticle velocity
$|\nabla_\k \Delta_\k|$, and the cutoff $\Lambda$ is of order
$\Delta_0$.
The integrals over $E$ and $\theta$ are easily done and
\[
g_{pq}(\omega) = \frac{-\alpha}{2^{p+q}}
[ \omega F_{p+q}(\omega) P^0_{pq} \hat \tau_0 + F_{p+q+1}(\omega)
(P^3_{pq} \hat \tau_3 + P^1_{pq} \hat \tau_1) ]
\]
where $P^j_{pq}$ are constants given by the angular integrations, and
\begin{eqnarray}
F_a(\omega) &=& \int_0^\Lambda EdE \frac{E^a}{E^2-\omega^2}\nonumber .
\end{eqnarray}
The constants $P^j_{pq}$ vanish for $j=1,3$ when $p+q = \mathrm{even}$
and vanish for $j=0$ when $p+q = \mathrm{odd}$.  The first few nonzero
elements are
\begin{eqnarray*}
 P^0_{00} &=& 1 \\
 P^1_{10} &=& -P^1_{01} = \frac{1}{2\Delta_0}\\
 P^3_{10} &=& P^3_{01} =  -\frac{1}{4t}\\
 P^0_{11} &=& P^0_{11} = -\frac{1}{2\Delta_0^2} + \frac{1}{8t^2}
\end{eqnarray*}
Only even moments of $F_a(\omega)$ are needed:
\begin{eqnarray*}
F_{2n} &=& \sum_{j=0}^{n-1} \frac{\omega^{2j}\Lambda^{2(n-j)}}{2(n-j)} +
\frac {\omega^{2n}}{2} \ln \frac{\Lambda^2}{-\omega^2}
\end{eqnarray*}
Since we are interested in the leading order behavior of $G(\R,\omega)$
we note that for small $\omega$ ,
\begin{eqnarray*}
F_0(\omega) &\rightarrow &  \frac {1}{2}
   \ln \frac{\Lambda^2}{-\omega^2} \\
F_{2n} (\omega) &\rightarrow &  \frac{\Lambda^{2n}}{2n}
\end{eqnarray*}

For $\R = (2m,2n)$, the leading order contribution to $G(\R,\omega)$ comes
from the single term in the expansion containing $g_{00}$.  To
second order in $\omega$:
\begin{equation}
G(\R,\omega) =
-(-1)^{n+m} \frac{\alpha \omega}{2}\ln \frac{\Lambda^2}{-\omega^2} \hat\tau_0
+ \omega C_0(\R) \hat \tau_0,
\end{equation}
where $C_0(\R)$ is real, and is the sum of several terms.  The largest
term contributing to $C(\R)$ is of order
\[
\frac{\alpha|\omega|}{16(m+n)}\left (
\frac{\Lambda}{\Delta_0}\right)^{2(m+n)}
\]
from which we estimate a range of validity
\[
|\omega| \lesssim \Lambda e^{-\frac{(\Lambda/\Delta_0)^{2(m+n)}}{ 16(m+n)}}.
\]
For other $\R$, there is no single dominant term in the expansion for
the Green's function, and the leading order behavior comes from the sum
over a large number of real nondivergent terms.  For our purposes,
it is sufficient to note that when
$\R = (2m+1,2n+1)$, the sums take the form
\begin{equation}
G(\R,\omega) = \omega C_0(\R) \hat \tau_0  ,
\end{equation}
and when $\R = (2m+1,2n)$ or $(2m,2n+1)$
\begin{equation}
G(\R,\omega) = C_1(\R) \hat \tau_1 + C_3(\R) \hat \tau_3,
\end{equation}
where $C_0(\R)$, $C_1(\R)$, and $C_3(\R)$ are real constants.

\end{document}